\documentclass[aps,prl,twocolumn,showpacs,superscriptaddress]{revtex4-1}
\usepackage{amsmath,amssymb}
\usepackage{graphicx}
\usepackage{subfigure}
\usepackage{verbatim}
\usepackage{amsfonts}
\usepackage{dcolumn}
\usepackage{bm}
\usepackage{ulem}
\usepackage{color}
\usepackage[colorlinks,citecolor=blue]{hyperref}

\usepackage{mathrsfs}

\newcommand{\beq}{\begin{eqnarray} }
	\newcommand{\eeq}{\end{eqnarray} }
\newcommand{\Beq}{\begin{eqnarray*} }
	\newcommand{\Eeq}{\end{eqnarray*} }
\newcommand{\Bmat}{\left(\begin{matrix}}
	\newcommand{\Emat}{\end{matrix}\right)}

\begin{document}
	
\title{Universal Rule for Topological Hopf Term via Dirac-Spin Coupling}

\author{Yan-Guang Yue}
\affiliation{Department of Physics and State Key Laboratory of Surface Physics, Fudan University, Shanghai 200433, P.R. China}

\author{Shuai Yang}
\affiliation{Department of Physics and State Key Laboratory of Surface Physics, Fudan University, Shanghai 200433,  P.R. China}

\author{Zheng-Xin Liu}
\affiliation{Department of Physics, Renmin University of China, Beijing 100872, China}

\author{Yan Chen}
\email{yanchen99@fudan.edu.cn}
\affiliation{Department of Physics and State Key Laboratory of Surface Physics, Fudan University, Shanghai 200433, P.R. China}
\affiliation{Shanghai Branch, Hefei National Laboratory, Shanghai 201315, P.R. China}

\begin{abstract}	
It is known that the topological Hopf term in two-dimensional (2D) spin systems can be derived by coupling to massless Dirac fermions. We establish a universal rule governing the generation of Hopf terms in 2D  quantum spin systems coupled to Dirac fermions. The key insight identifies the Hopf coefficient as the oriented volume in the $\mathfrak{su}(2)$ Lie algebra space formed by Dirac cone matrix bases. This geometric interpretation allows direct determination of Hopf term without path integral computations. Applying this framework, we demonstrate nontrivial Hopf term in tailored checkerboard lattice models and recover known results in graphene-based systems.
\end{abstract}

\maketitle
\textit{Introduction.—}Over the past few decades, the study of topological states of matter has emerged as a central topic in condensed matter physics and quantum field theory. Many topological phases can be characterized by topological terms in their path integrals—these terms, typically independent of the spacetime metric in continuum quantum field theory, capture the global topological properties of the system. For instance, both integer and fractional quantum Hall liquids \cite{ZHK_CS_FQH, LopezFradkin_CS_FQH} as well as chiral spin liquids \cite{KL_CSL,WenWilczekZee_CSL,Wen_CSL} can be described by Chern-Simons terms in effective gauge theories. Additionally, the topological $\theta$-term plays a crucial role in constructing the theory of topological insulators \cite{PRB781}. The Haldane phase can be understood by introducing a $\theta$-term into  (1+1)-D $SO(3)$ nonlinear sigma model (NLSM) \cite{HaldanePLA1983, HaldanePRL1983}, where the topological $\theta$-term is quantized to integer multiples of 2$\pi$ when the spacetime manifold is closed. If the system has a boundary, the $\theta$-term can be identified with the Berry phase of spin-½ edge states \cite{Ng1994}.

Topological terms exhibit unique effects in gapless systems. For example, the spin-½ antiferromagnetic Heisenberg chain, which includes a $\pi$-quantized $\theta$-term or Wess-Zumino-Witten term, obeys the Lieb-Schultz-Mattis theorem and thus remains gapless \cite{WZW,LSM}. In gapped systems, while some exhibit intrinsic topological order (characterized by fractionalized excitations and chiral edge states), there exists a class of systems that, despite lacking intrinsic topological order, still host edge states protected by symmetries—these are known as symmetry-protected topological (SPT) states \cite{GuWen2009, ChenGuLiuWen2011, ChenGuLiuWenSCI}.

The spin-1 Haldane phase mentioned earlier is a 1D SPT phase. 1D SPT phases can be described by projective representations and classified by the second group cohomology of the symmetry group \cite{ChenGuWen2011_1Dfull,ChenGuWen2011_1D}. The Haldane phase, described by (1+1)-D $SO(3)$ NLSM with a $\theta$-term, can be generalized to higher dimensions, leading to bosonic SPT phases described by (d+1)-D $SO(d+2)$ NLSM with higher-order $\theta$-terms \cite{PRB9113}. This essentially arises from the mapping of a compactified (d+1)-D spacetime manifold $S^{d+1}$ to the symmetric space $S^{d+1}$ of the $SO(d+2)$ group, classified by the homotopy group $\pi_{d+1}(S^{d+1})=\mathbb{Z}$.

Another important generalization is (2+1)-D $SO(3)$ NLSM with a Hopf term, originating from the Hopf map from the spacetime manifold $S^3$ to the symmetric space $S^2$, classified by the homotopy group $\pi_3(S^2)=\mathbb{Z}$. The Hopf term alters the statistical properties of skyrmions \cite{WilczekZee, WuZee}. Consequently, a spin system with a Hopf term in its Lagrangian and a gapped ground state may exhibit either intrinsic topological order \cite{WenNiu90, Wen1990} or symmetry-protected topological order \cite{LiuWen}.

One well-established approach to generating the Hopf term in spin systems is to couple spins to massless Dirac fermions \cite{Abanov,AbanovWiegmann}. Upon integrating out the fermions, a Hopf term can emerge in the effective action of the spin field. However, this process typically involves technically demanding path integral calculations and offers little predictive power—it is generally unclear in advance which specific Dirac-spin coupling will lead to a nontrivial Hopf term. This raises a fundamental question: Is there a general principle that allows us to predict the emergence and magnitude of the Hopf term directly from the algebraic structure of Dirac-spin coupling? Such a principle would significantly streamline the analysis of topological terms in spin-fermion systems and offer practical guidance in model construction.

In this work, we propose a universal and geometrically intuitive rule that relates the Hopf term to the oriented volume formed by a set of $\mathfrak{su}(2)$ matrices defining the Dirac-spin coupling. This approach enables direct evaluation of the topological response without resorting to path integrals. We outline the key steps in the main text and provide a full derivation in the supplemental material. The rule is validated across several representative models, including a lattice model we construct on the checkerboard geometry, incorporating staggered chemical potentials and extended hopping. In this setting, two Dirac cones of identical chirality appear in the Brillouin zone, and coupling them to a spin field with short-range antiferromagnetic correlations yields a nontrivial Hopf term.
\begin{figure}[htbp]
	\centering
	\includegraphics[width=0.9\linewidth]{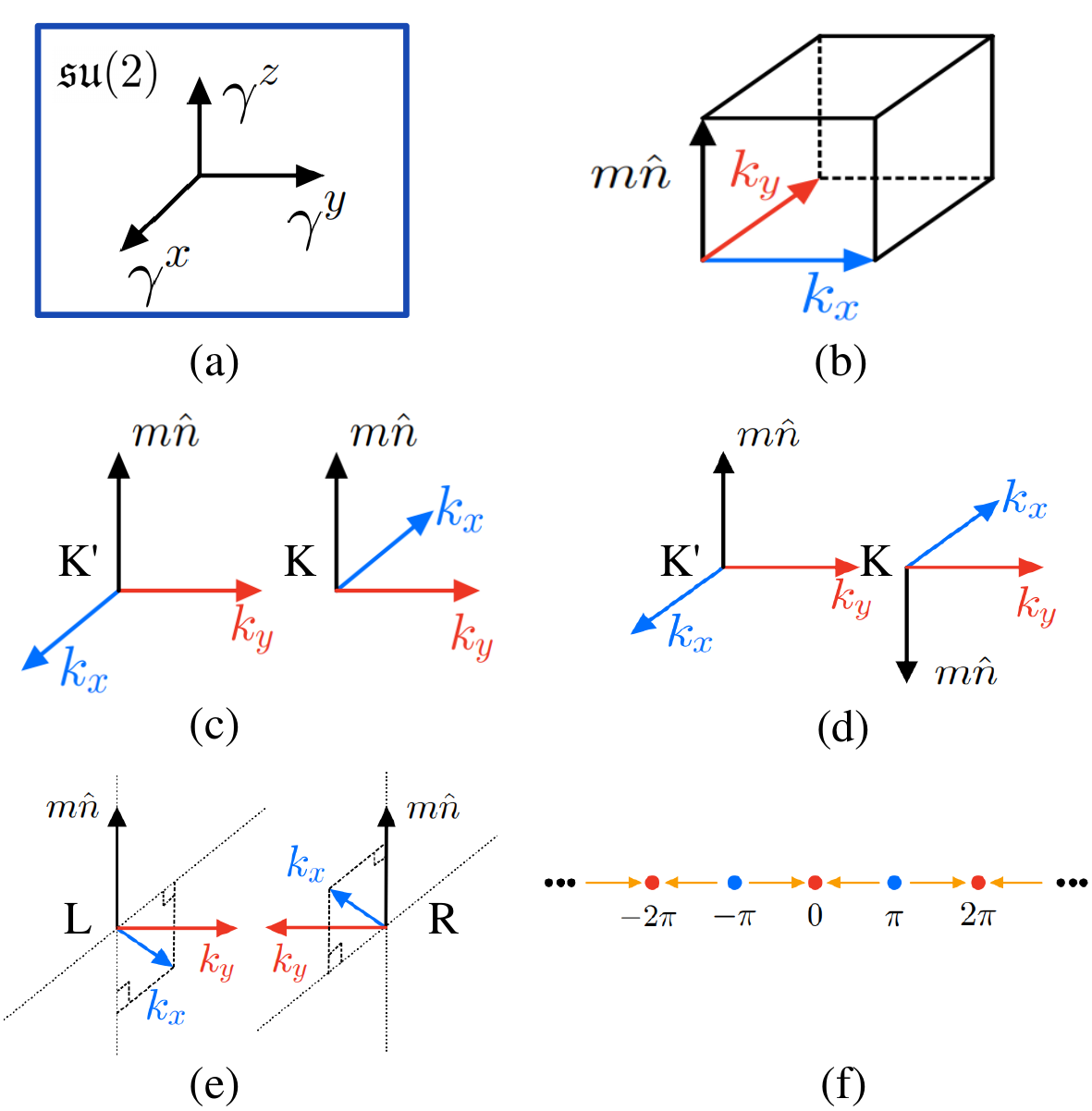}
	\caption[2band]{Normalized matrix basis for some models. (a) Pauli matrices serve as bases for the $\mathfrak{su}(2)$ Lie algebra  of the  $SU(2)$ group. (b)The oriented volume of a set of matrix basis tensors. The red arrows, blue arrows, and black arrows in the last figures represent the matrices that $k_x,k_y,m\hat{n}$ follow, respectively. (c,d) Graphene model and the four-band model. (e) The two-cone model. (f) $\theta$ at a non-fixed point under RG will flow to a stable fixed point. The red dots are stable fixed points, and the blue ones are unstable fixed points.}
	\label{bases}
\end{figure}

Several 2D lattice models host Dirac-cone band structures in their BZs. For the $j$-th Dirac cone in the target model, the low-energy effective Hamiltonian of massless Dirac fermions can be generally written as $v_x k_x \gamma_j^{\prime x}+v_y k_y \gamma_j^{\prime y}$, where $v_x$ and $v_y$ are the Fermi velocities along the $x-$  and $y-$directions, which can be anisotropic. When a spin field $\boldsymbol{n}(x)$ is introduced on the lattice, it couples to the Dirac fermions via a coupling term whose low-energy effective Hamiltonian is given by $m\hat{n}\, \tilde{\gamma}_j^{\prime z}$. Thus, the low-energy effective Hamiltonian for the $j$-th Dirac cone becomes
\begin{equation}
\begin{aligned}
     H_{\boldsymbol{K}_j+\boldsymbol{k}} = v_x k_x\, \gamma^{\prime x}_j + v_y k_y\, \gamma^{\prime y}_j + m\hat{n}\, \tilde{\gamma}_j^{\prime z},
\end{aligned}
\end{equation}
Here, $\boldsymbol{k}=\left(k_x, k_y\right)$ denotes the small momentum measured relative to the Dirac point at $\boldsymbol{K}_j$; $m$ represents the magnitude of the spin; $\hat{n}=\boldsymbol{n} \cdot \boldsymbol{\sigma}$; and $\gamma_j^{\prime x}, \gamma_j^{\prime y}$, and $\tilde{\gamma}_j^{\prime z}$ are the corresponding normalized $2 \times 2$ Hermitian matrices that serve as the basis of the low-energy Dirac–spin coupled Hamiltonian. The precise meaning of ``normalization" will be explained later. Upon integrating out the $j$-th Dirac fermion, a Hopf term may be generated in the effective action. Identifying a suitable model and coupling that give rise to a Hopf term is highly nontrivial. In fact, the Hopf term was only recently realized for the first time in two-dimensional lattice models \cite{PRB107}.

\textit{Universal rule.—}We now present the universal rule for determining the Hopf term generated by Dirac fermions coupled to a spin field. For the $j$-th Dirac cone in the target model, its normalized matrix triplet-$(\gamma'^{x}_j, \gamma'^{y}_j, \tilde{\gamma}'^{z}_j)$ can be regarded as vectors in the su(2) Lie algebra associated with the $\mathfrak{su}(2)$ group. Within this algebraic space, there exists a natural bi-invariant metric known as the Killing–Cartan metric, defined as:
   \begin{equation}
    \begin{aligned}
    \langle X, Y \rangle = \frac{1}{2}\operatorname{Tr}(X Y), \quad X,Y \in \mathfrak{su}(2).
   \end{aligned}
   \end{equation} 
Under this inner product, the standard Pauli matrices $\left(\gamma^x, \gamma^y, \gamma^z\right)$ form an orthonormal basis of the $\mathfrak{s u}(2)$ vector space  vector space [Fig.~\ref{bases}(a)]. In such a space, one can define the oriented volume of three matrices as:  
   \begin{equation}
    \begin{aligned}
    V = -\frac{i}{2}\operatorname{Tr}\Bigl(XYZ\Bigr),\quad X,Y,Z \in \mathfrak{su}(2).
    \end{aligned}
   \end{equation}
This oriented volume captures the  chirality of the basis set. Applied to the Dirac cone, the three normalized matrix bases $(\gamma'^{x}_j, \gamma'^{y}_j, \tilde{\gamma}'^{z}_j)$ span a volume in $\mathfrak{s u}(2)$ space given by: $V_j = -\frac{i}{2}\operatorname{Tr}\Bigl(\gamma'^{x}_j\gamma'^{y}_j\tilde{\gamma}'^{z}_j\Bigr)$ [Fig.~\ref{bases}(b)]. This geometric interpretation reveals a key physical insight: the emergence and magnitude of the Hopf term are determined by the mutual orientation of the normalized matrix triplet.  Thus, the Hopf term becomes a purely geometric quantity—determined entirely by the internal structure of the Dirac cone—eliminating the need for path integral calculations. The total Hopf term of a system with multiple Dirac cones is simply the sum over all cones:
\begin{equation}
\begin{aligned}
S_{\text{tHopf}} = \left(\sum_j V_j\right) \,\pi\, H(\boldsymbol{n}),\label{rule}
\end{aligned}
\end{equation}
where $H(\boldsymbol{n})=-\frac{1}{4 \pi^2} \int d^3 x \epsilon^{\mu \nu \rho} a_\mu \partial_\nu a_\rho$ denotes the Hopf invariant corresponding to the mapping $S^3 \to S^2$ defined by the spin field $\boldsymbol{n}$. This rule implies that the Hopf term contribution from a single Dirac cone always lies within the range $[-\pi, \pi]$, with the sign set by the cone's chirality and the absolute value given by the volume of the corresponding matrix triple. The boundary value $\pm \pi$ is achieved when the three normalized matrices are mutually orthogonal. 
   
\textit{Proof of the rule.—}We derive the universal rule by analyzing the low-energy effective action of a spin-coupled Dirac system. Starting from an effective Hamiltonian near a Dirac point: $H_k=v_x k_x \gamma^{\prime x}+v_y k_y \gamma^{\prime y}+m \hat{n} \tilde{\gamma}^{\prime z}$. We prove that the magnitude of the Fermi velocity does not affect the size of the Hopf term. Additionally, we show that the momentum space can always be rotated such that the re-obtained $\gamma^{\prime x}$ and $\gamma^{\prime y}$ are orthogonal with respect to the Killing-Cartan metric. Consequently, we can always find $\gamma^{\prime z}$ in the Lie algebra space, along with $\gamma^{\prime x}$ and $\gamma^{\prime y}$, which together form an orthonormal basis and share the same chirality as the Pauli matrices. The $\cos \theta= \langle \tilde{\gamma}^{\prime z}, \gamma^{\prime z} \rangle$ is the angle between $\tilde{\gamma}^{\prime z}$ and $\gamma^{\prime z}$. We obtain the effective action:
$S_{\mathrm{eff}}=-i \operatorname{Tr} \ln D,$
where $D=i \gamma^\mu \partial_\mu+m\left(\cos \theta \hat{n}+\sin \theta \gamma^1 \hat{n}\right)$, and the gamma matrices are defined as: $
\gamma^0=-\gamma^{\prime z}, \quad \gamma^1=\gamma^0 \gamma^{\prime x}, \quad \gamma^2=\gamma^0 \gamma^{\prime y}$ which satisfy the Clifford algebra $\left\{\gamma^\mu, \gamma^\nu\right\}=2 g^{\mu \nu}$ with the metric signature $g^{\mu \nu}=(1,-1,-1)$. Expanding the variation of $S_{\text {eff }}$ in powers of $m$, the topological Hopf term emerges at third order:
\begin{equation}
\begin{aligned}
\delta S_{\text {topo }}=-\frac{\cos \theta}{32 \pi} \int d^3 x \epsilon^{\mu \nu \rho} \operatorname{Tr}\left(\delta \hat{n} \hat{n} \partial_\mu \hat{n} \partial_\nu \hat{n} \partial_\rho \hat{n}\right)
\end{aligned}
\end{equation}
which simplifies to: $\delta S_{\text {topo }}=-\frac{\cos \theta}{2 \pi} \int d^3 x \epsilon^{\mu \nu \rho} \delta a_\mu \partial_\nu a_\rho$
upon substituting $\hat{n}=2 z z^{\dagger}-1$, where $a_\mu=z^{\dagger}\left(-i \partial_\mu\right) z$, with $z^t=\left(z_1, z_2\right)$ being a complex vector of unit modulus, satisfying $z^{\dagger} z=1$. The total Hopf action becomes: $S_{\text {topo }}=\cos \theta \pi H(\boldsymbol{n})$.
Crucially, the prefactor $\cos \theta$ encodes the oriented volume $V=-\frac{i}{2} \operatorname{Tr}\left(\gamma^{\prime x} \gamma^{\prime y} \tilde{\gamma}^{\prime z}\right)$ of the normalized basis $\left(\gamma^{\prime x}, \gamma^{\prime y}, \tilde{\gamma}^{\prime z}\right)$ for the Dirac cone. For systems with multiple cones, the total Hopf term is the sum of contributions from all cones: $S_{\mathrm{tHopf}}=\left(\sum_j V_j\right) \pi H(\boldsymbol{n})$, thereby establishing the universality of the rule.

\begin{figure}
    \centering
\includegraphics[width=0.9\linewidth]{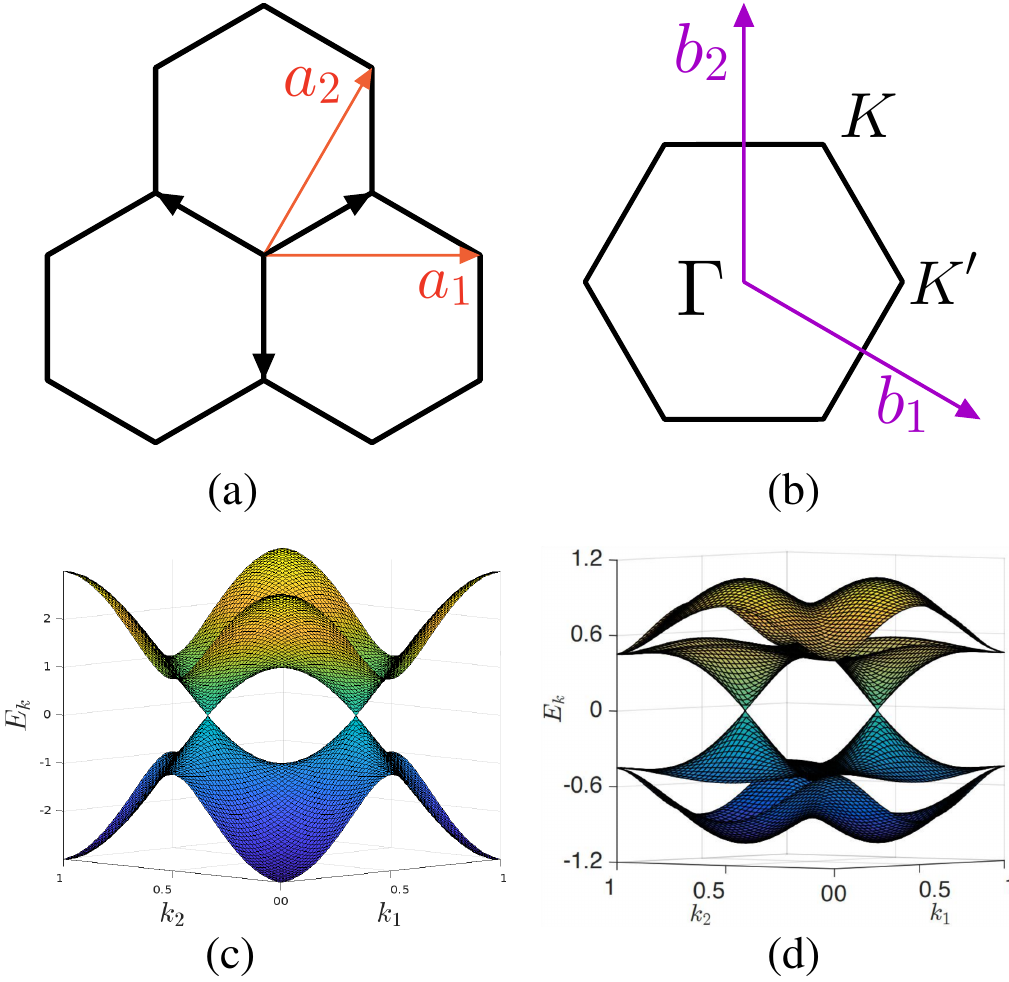}
    \caption{(a) Honeycomb lattice and lattice basis vectors. (b) Reciprocal lattice vectors and Brillouin zone. (c) Band structure of the graphene tight-binding model. (d) Band structure of the four-band model.}
    \label{four band}
\end{figure}

\textit{Validation of the rule.—}To validate the proposed universal rule for determining the emergence of Hopf terms, we apply it to two representative lattice models with well-established results. The first is the graphene model \cite{PRB107} on a honeycomb lattice \cite{PRB29,Graphene}, where electrons hop between nearest neighbors. The tight-binding Hamiltonian 
$H_0 = \sum_{\langle i,j \rangle,\sigma=\uparrow,\downarrow} \left( t\, c_{i\sigma}^\dagger c_{j\sigma} + \mathrm{h.c.} \right)$, yields two Dirac cones located at the high-symmetry points $\boldsymbol{K}^{\prime}$ and $\boldsymbol{K}$ in the BZ [Fig.~\ref{four band}(b)(c)]. The corresponding low-energy Hamiltonians near these points take the form: $H_{\boldsymbol{K}^{\prime}+\boldsymbol{k}}=v\left(k_x \gamma^x+k_y \gamma^y\right)$ and $H_{\boldsymbol{K}+\boldsymbol{k}}=v\left(-k_x \gamma^x+k_y \gamma^y\right)$, with $v$ the Fermi velocity. We assign a spin degree of freedom to each lattice site, characterized by a local angular momentum expectation value $\langle S_i \rangle = \boldsymbol{n}_i$, where $\boldsymbol{n}_i$ is assumed to be a smooth function of the site index $i$. The electron-spin interaction is described by the Hamiltonian \begin{equation}
\begin{aligned}
H_{\text{cp}}  =\sum_i f(i) m C_i^{\dagger} \boldsymbol{\sigma} C_i \cdot \boldsymbol{n}_i =\sum_{k, q} m C_k^{\dagger} \gamma^z \boldsymbol{\sigma} C_q \cdot \boldsymbol{n}_{k-q},\label{acp}
\end{aligned}
\end{equation}where $C_i^{\dagger} = (c_{i \uparrow}^{\dagger}, c_{i \downarrow}^{\dagger})$, $C_k = \sum_i C_i, e^{i k \cdot r_i}$, and $f(i) = +1$ ($-1$) for sites on the $A$($B$) sublattice. Due to the sublattice-dependent sign of the coupling, electrons on the $A$ and $B$ sublattices experience opposite effective magnetic moments, indicating that the spin texture exhibits short-range antiferromagnetic correlation. Projecting onto the low-energy subspace, the effective Hamiltonians read $H_{\boldsymbol{K}^{\prime}+\boldsymbol{k}}=v k_x {\gamma^x}+v k_y {\gamma^y}+m \hat{n} {\gamma^z}$ and $H_{\boldsymbol{K}+\boldsymbol{k}}=$ $v k_x\left({-\gamma^x}\right)+v k_y{\gamma^y}+m \hat{n} \gamma^z$. The normalized matrix bases for the two cones are given by $\left(\bm{\textcolor{blue}{\gamma^x}}, \bm{\textcolor{red}{\gamma^y}}, \bm{\gamma^z}\right)$ and $\left(\bm{\textcolor{blue}{-\gamma^x}}, \bm{\textcolor{red}{\gamma^y}}, \bm{\gamma^z}\right)$ [Fig.~\ref{bases}(c)], enclose volumes in the Lie algebra space that exactly cancel: 
\begin{equation}
\begin{aligned}
V_{\boldsymbol{K}^{\prime}}+V_{\boldsymbol{K}}=\frac{-i}{2} \operatorname{Tr}\left(\bm{\textcolor{blue}{\gamma^x}}\bm{\textcolor{red}{\gamma^y} }\bm{\gamma^z}\right)+\frac{-i}{2} \operatorname{Tr}\left(\bm{\textcolor{blue}{-\gamma^x}} \bm{\textcolor{red}{\gamma^y} }\bm{\gamma^z}\right)=0.\notag
\end{aligned}
\end{equation}
This confirms the absence of the Hopf term, consistent with previous studies.

The second example is the four-band model \cite{PRB107,WuCJ1,WuCJ2}, which realizes the Hopf term for the first time in 2D lattice models. This model studies a tight-binding system of $p_x$ and $p_y$ orbitals on a honeycomb lattice. Hiding spin indexes and introducing the eigen bases of the orbital angular momentum operator $l_z$, i.e., $C_{i \pm }=\left(C_{i x a} \pm i C_{i y}\right) / \sqrt{2}$. Ignoring the
spin-orbit coupling, then the tight binding model with nearest neighbor hopping reads $H_0=\sum_{\langle i,j \rangle}( h_{++}^{ij} C_{i+}^\dagger C_{j+} + h_{+-}^{ij} C_{i+}^\dagger C_{j-} + h_{-+}^{ij} C_{i-}^\dagger C_{j+} 
+ h_{--}^{ij} C_{i-}^\dagger C_{j-} + \mathrm{h.c.})$, where $h_{ \pm \pm}^{i j}$ are bond-dependent hopping constants. The resulting band structure features four bands, with Dirac cones formed between the middle two bands at $\boldsymbol{K}^{\prime}$ and $\boldsymbol{K}$ [Fig.~\ref{four band}(b)(d)]. The corresponding low-energy Hamiltonian near these points take the form: $H_{\boldsymbol{K}^{\prime}+\boldsymbol{k}}=$ $v k_x \gamma^x+v k_y \gamma^y$ and $H_{\boldsymbol{K}+\boldsymbol{k}}=v k_x\left(-\gamma^x\right)+v k_y \gamma^y$. Again, the spin texture $\boldsymbol{n}_i$ is then introduced that couples with opposite sign to the $l_z= \pm 1$ orbitals: $H_{\text{cp}} = \sum_i m \left( C_{i+}^\dagger \boldsymbol{\sigma}\, C_{i+} \cdot \boldsymbol{n}_i - C_{i-}^\dagger \boldsymbol{\sigma}\, C_{i-} \cdot \boldsymbol{n}_i \right)$. Projecting onto the low-energy subspace, the effective
Hamiltonians at the two Dirac points are $H_{\boldsymbol{K}^{\prime}+\boldsymbol{k}} = vk_x {\gamma^x} + vk_y{\gamma^y} + m\hat{n}{\gamma^z},
H_{\boldsymbol{K}+\boldsymbol{k}}= vk_x\, {(-\gamma^x)} + vk_y{\gamma^y} + m\hat{n}{(-\gamma^z)}$. The normalized matrix bases for the two cones are given by $\left(\bm{\textcolor{blue}{\gamma^x}}, \bm{\textcolor{red}{\gamma^y}}, \bm{\gamma^z}\right)$ and $\left(\bm{\textcolor{blue}{-\gamma^x}}, \bm{\textcolor{red}{\gamma^y}}, \bm{\gamma^z}\right)$ [Fig. \ref{bases}(d)]. According to the rule, the total oriented volume evaluates to 
\begin{equation}
\begin{aligned}
V_{\boldsymbol{K}^{\prime}}+V_{\boldsymbol{K}}=\frac{-i}{2} \operatorname{Tr}\left(\bm{\textcolor{blue}{\gamma^x} }\bm{\textcolor{red}{\gamma^y}} \bm{\gamma^z}\right)+\frac{-i}{2} \operatorname{Tr}\left(\bm{\textcolor{blue}{-\gamma^x} }\bm{\textcolor{red}{\gamma^y}}\bm{\left(-\gamma^z\right)}\right)=2,\notag
\end{aligned}
\end{equation} in agreement with the known quantized Hopf term of $2 \pi$. While this model demonstrates a nontrivial Hopf term, it does so at the cost of additional orbital degrees of freedom. Next, we seek a more minimal realization on a checkerboard lattice featuring only two bands.

\textit{Two-cone model.—}To generate a nontrivial Hopf term via Dirac-spin coupling, the system must host at least two Dirac cones. According to the universal rule, a single Dirac cone contributes at most a Hopf term of magnitude $\pi$, which corresponds to an unstable fixed point under renormalization group (RG) flow and is likely to flow to zero. As a result, the system may remain gapless \cite{IJMPA1990,PRLxu}. Moreover, the fermion doubling theorem implies that Dirac cones typically appear in pairs with opposite chirality in systems preserving time-reversal symmetry \cite{DBF}. Therefore, realizing a nontrivial Hopf term requires constructing a model where two Dirac cones share the same chirality, necessitating the breaking of time-reversal symmetry. A suitable starting point is the two-band model on the checkerboard lattice featuring a quadratic band crossing point (QBCP) at the BZ, as discussed in Ref.~\onlinecite{PRL106}.
\begin{figure}[htbp]
	\centering
 \includegraphics[width=0.9\linewidth]{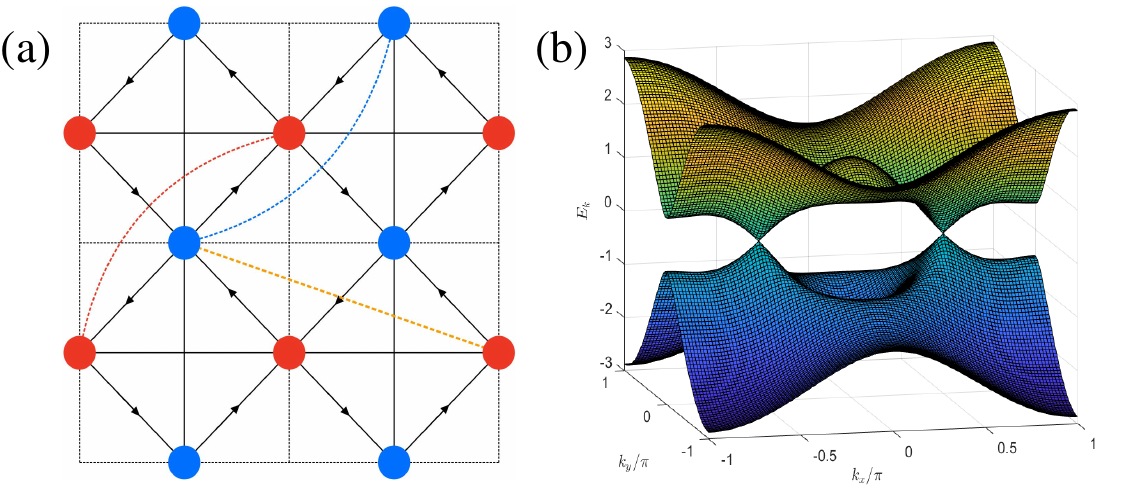}
	\caption[2band]{(a) The checkerboard lattice. Arrows and black solid lines represent the nearest-neighbor and next-nearest-neighbor hoppings, respectively. The red and blue dashed curves indicate third-nearest-neighbor hoppings, while the yellow curve represents the fourth-nearest-neighbor hopping.
(b) Two Dirac cones in the BZ. }
\label{the fourth}
\end{figure} We consider a modified version of this model that includes hopping terms up to the fourth nearest neighbor and a staggered chemical potential. The Hamiltonian reads:
\beq
H_0&=& -t e^{i\phi_{ij}}\sum_{\langle i,j\rangle} (c_i^\dag c_j + \mathrm{h.c.}) - t^{\prime}\sum_{\langle\langle i,j\rangle\rangle} (c_i^\dag c_j + \mathrm{h.c.}) \notag\\
&&- t^{\prime\prime}\sum_{\langle\langle\langle i,j\rangle\rangle\rangle} (c_i^\dag c_j + \mathrm{h.c.})- t^{\prime\prime\prime}\sum_{\langle\langle\langle\langle i,j\rangle\rangle\rangle\rangle} (c_i^\dag c_j + \mathrm{h.c.})\notag\\
&&-M\sum_{i}f(i)c_i^\dag c_i,
\eeq
where $\phi_{i j}$ introduces a complex phase on the nearest-neighbor hopping, breaking time-reversal symmetry. The site-dependent function $f(i)=+1(-1)$ if site $i$ belongs to the $\mathrm{A}(\mathrm{B})$ sublattice, resulting in a staggered chemical potential. The four hopping terms correspond to 1st through 4th nearest neighbors, respectively [Fig.~\ref{the fourth}(a)].
Transforming to momentum space, the Hamiltonian becomes a $2 \times 2$ matrix in the sublattice basis:
\beq
\mathcal{H}_0&=&-[(t'_1+t'_2)(\cos k_x+\cos k_y)+4t''\cos k_x\cos k_y]\mathbb{I}\notag\\&&-\big[4t\cos\phi(\cos\frac{k_x}{2}\cos\frac{k_y}{2})\notag\\&&-4t^{\prime\prime\prime}(\cos{\frac{k_{x}}{2}}\cos{\frac{3k_{y}}{2}}+\cos{\frac{k_{y}}{2}}\cos{\frac{3k_{x}}{2}})\big]\gamma^x\notag\\&&-4t\sin\phi(\sin\frac{k_x}{2}\sin\frac{k_y}{2})\gamma^y\notag\\ &&-\big[(t'_1-t'_2)(\cos k_x-\cos k_y)+M\big]\gamma^z,\label{ham}
\eeq
geometry. By choosing $t_1^{\prime}=-t_2^{\prime}=t /(2 \sqrt{2}), t^{\prime \prime}=0, t^{\prime \prime \prime}=t / 4$, and $M=2 t^{\prime}$, the band structure hosts two Dirac cones with the same chirality located at $\boldsymbol{R}=$ $(\pi / 2,0)$ and $\boldsymbol{L}=(-\pi / 2,0)$ [Fig.~\ref{the fourth}(b)]. This configuration results from the breaking of time-reversal symmetry by the phase $\phi=\pi / 2$, which splits the original QBCP into two Dirac points with the same chirality. We refer to this setup as the ``two-cone model".

To analyze the topological response, we expand the Hamiltonian near the two Dirac points and define the valley spinors $\Psi_{\boldsymbol{R}(\boldsymbol{L})}^{\dagger}(\mathbf{k})=\left(\psi_{\boldsymbol{R}(\boldsymbol{L}), A}^{\dagger}, \psi_{\boldsymbol{R}(\boldsymbol{L}), B}^{\dagger}\right)$. The lowenergy effective Hamiltonians near the two valleys are:
\begin{equation}
\begin{aligned}
\left\{
    \begin{array}{ll}
        H_{\boldsymbol{R}+\boldsymbol{k}} &=v k_x \frac{\sqrt{2}}{2}(\gamma^{x} + \gamma^{z}) + vk_y (-\gamma^{y})\\[2.5mm]
H_{\boldsymbol{L}+\boldsymbol{k}} &= vk_x \frac{-\sqrt{2}}{2}(\gamma^{x} + \gamma^{z}) +vk_y \gamma^{ y},
    \end{array}
    \right.
\end{aligned}
\end{equation}
We now introduce a spin field $\mathbf{n}_i$ coupled to the fermions via a staggered interaction as in Eq.~(\ref{acp}). Upon projection to the low-energy subspace, the spin coupling yields effective mass terms:
\begin{equation}
\begin{aligned}
\left\{
    \begin{array}{ll}
        H_{\boldsymbol{R}+\boldsymbol{k}} &= v k_x {\frac{\sqrt{2}}{2}(\gamma^{x} + \gamma^{z})} 
+ v k_y {(-\gamma^y)} + m \hat{n} {\gamma^z}\\[2.5mm]
H_{\boldsymbol{L}+\boldsymbol{k}} &= v k_x {\frac{-\sqrt{2}}{2}(\gamma^{x} + \gamma^{z})}
+ v k_y {\gamma^y} + m \hat{n} {\gamma^z}.
    \end{array}
    \right.
\end{aligned}
\end{equation}
The three matrices involved in each Dirac cone-namely, $\gamma^{\prime x}=\frac{\sqrt{2}}{2}\left(\gamma^x+\gamma^z\right)$, $\gamma^{\prime y}=-\gamma^y$, and $\tilde{\gamma}^{\prime z}=\gamma^z$ for the $\boldsymbol{R}$ cone (with opposite signs for $\boldsymbol{L}$ )—define the geometry relevant to the Hopf term [Fig.~\ref{bases}(e)]. Applying the universal formula Eq.~(\ref{rule}), we obtain the total Hopf coefficient:
 \begin{equation}
\begin{aligned}
S_{\rm tHopf} =& \Big(V_{\boldsymbol{R}} + V_{\boldsymbol{L}}\Big) \,\pi\, H(\boldsymbol{n}) \\
=& \Big(\frac{-i}{2} {\rm Tr} \bm{\textcolor{blue}{\frac{\sqrt{2}}{2}(\gamma^{x} + \gamma^{z})}} \bm{\textcolor{red}{(-\gamma^y)}} \bm{\gamma^{z}}\\
&+\frac{-i}{2} {\rm Tr}\bm{\textcolor{blue}{\frac{-\sqrt{2}}{2}(\gamma^{x} + \gamma^{z})}}\bm{\textcolor{red}{\gamma^y}}\bm{ \gamma^{z} }
 \Big) 
\,\pi\, H(\boldsymbol{n}) \\
= &-\sqrt{2} \,\pi\, H(\boldsymbol{n}).
\end{aligned}
\end{equation}
In our analysis, the effective $\theta$ term arising from Dirac-spin coupling is not an independent parameter but is instead subject to the RG flow. Nonperturbative arguments show that, when the dynamic term becomes unimportant in the strong coupling regime, the only remaining contribution is the topological $\theta$ term. In such cases, RG analyses \cite{PRLxu,PRB107} indicate that the $\theta$ parameter flows toward the nearest fixed point corresponding to $\theta=2 \pi N$, with $N \in \mathbb{Z}$ [Fig.~\ref{bases}(b)]. Fixed points at odd multiples of $\pi$ are unstable, so any bare value of $\theta$ that differs slightly from a quantized value will be driven by RG to the stable fixed point (e.g., a bare $\theta=-\sqrt{2} \pi$ will eventually flow to $-2 \pi$ ). This mechanism ensures that the low-energy physics of our two-cone model is governed by a quantized Hopf term, thereby stabilizing the corresponding SPT phase.

\textit{Conclusion.—}In this work, we propose a universal rule that determines the magnitude of the Hopf term directly from the internal geometric structure of Dirac–spin coupling—specifically, from the oriented volume formed by the normalized matrix basis in the su(2) Lie algebra. We have verified the consistency and generality of this rule across several representative models, including the graphene model, a four-band model, and a two-cone model constructed on a checkerboard lattice. Interpreted through geometric visualization, this method is not only intuitive but also computationally efficient and predictive, providing a powerful and novel tool for analyzing Hopf terms in topological spin systems.

\textit{Acknowledgment.—}We thank Kai Sun for the helpful discussions. This work is supported by the National Key Research and Development Program of China (Grant No. 2022YFA1402204) and the National Natural Science Foundation of China (Grant No. 12274086).

\bibliography{hopf}

\end{document}